# Modelling mobility and visualizing people's flow patterns in rural areas for future infrastructure development as a good transnational land-governance practice

Paula Botella[1], Paweł Gora[2], Martyna Sosnowska[3], Izabela Karsznia[3], Sara Carvajal Querol[4]

[1] School of Civil Engineering, Polytechnic University of Madrid
[2] Faculty of Mathematics, Informatics and Mechanics, University of Warsaw
[3] Faculty of Geography and Regional Studies, University of Warsaw
[4] Alianza por la Solidaridad NGO

**Abstract:** This paper summarizes a cross-border mobility study, origin-destination mobility modelling and visualization, conducted in support of the infrastructure development efforts of local authorities and NGOs on the area over the Kayanga-Geba River, at the border between Senegal and Guinea Bissau. It builds on the data collected through participatory mapping for the elaboration of the Cross-Border Land Management and Development Plans (Plans PAGET) aiming to harmonize the different national territorial management tools into a unique transnational tool through the consideration of border areas as a territorial unity. Despite a small amount of available mobility data, we were able to build a mobility model for the considered area, and implemented it in the Traffic Simulation Framework, which was later used to calculate origin-destination matrices for the studied regions in two cases: with and without a cross-border mobility. We analyzed the differences in the mobility patterns and visualized the mobility flows, deliberating on what may be the potential impacts of building a bridge in the study area. Our methodology is general and can be applied in similar studies on different areas. However, the quality of results may depend on the available data.

# 1. Introduction

Studying cross-border mobility becomes especially relevant when linked to governance, as it challenges the capacity of the current nation-states to respond to the needs (understood in this paper as socially and culturally defined, hence, allowing for a broad and variable interpretation depending on the context) of their citizens[1]. We base the work exposed in this paper on the hypothesis of transborder mobility[2] as a practice of necessity-driven social territorial construction[3] [16]: national state's boundaries are transgressed due to the lack of inner-access to the service infrastructure and social networks that sustain, and allow to improve, life; through this process, new territories are configured. Hence, cross-border mobility patterns invite us to reflect on alternatives to a form of governance limited to state's administrative borders, perceived, since movement is registered past these limits, as insufficient to guarantee inhabitants' well-being. Questions of belonging, ownership, competences and cross-country collaboration are then put on the spotlight.

---

[1] Questioning the good-will of governments and power institutions is something that remains outside of this research's objectives.
[2] In this paper *mobility* is used interchangeably with, and limited to, human physical mobility. Although the movement of nature and capital do also have a clear role in the configuration of territories, the study of these circuits goes beyond the purpose of this paper. Additionally, the relations between physical and cultural mobility, although implied, are not carefully examined.

The above challenge is of particular relevance in Africa, a continent where national borders were imposed; defined based on European political interests rather than sociolinguistic drivers, which translated into the separation of many ethnic groups [34]. Despite the division, traditional ethnic-based solidarity and connections have remained, becoming the main motivators of cross-border mobility, especially in those areas where *border asymmetries* in the access to services have been found [16].

The rural nature of most of these border areas should also be taken into account. From a development perspective, rural border areas are often beyond the service provision capacity of national governments [18]. Hence, the study of rural border areas has the potential to shed some light on the above mentioned questions on governance: traditional cross-border networks operating in these rural areas could be conceptualized as one of these governmental alternatives and complements to national administrations. By studying mobility patterns in rural cross-border areas, our goal is to contribute to a needs-based development, supporting the capacity of local transnational authorities to understand and respond to the main local challenges, recognizing and incorporating the value of traditional ties into their development strategies, and building their capacity to dialogue with mainstream, top-down nation-states frameworks.

In this study, an infrastructure demand analysis to justify the construction of a bridge within the study area is used as a motivation to reflect on the potential linkages between the study of mobility and improved transborder local governance. The main objective is to understand how mobility modelling and visualization, based on participatory data collection, can become a tool to support transnational, local and collaborative efforts to ensure the well-being of rural inhabitants living in border areas; those whose well-being builds on their capacity to access the resources of two countries. To this end, first, an introduction to the context and previous work leading to this study is made (Section 2), used to justify the purpose and main research questions behind this paper (Section 3). Following, a literature review of existing cross-border mobility studies, as well as the most suitable mobility modelling and visualization methodologies for the aimed work (Section 4). Next, an explanation of the methodology is included (Section 5) to end with the presentation and analysis of the obtained results (Section 6), conclusion and areas for future research (Section 7).

## 2. Socio-political context of the Senegal-Bissau border

The presented study took place at the border between Southern Senegal (an area known as Casamance) and Guinea-Bissau, it covered the PAGET South area as shown on Figure 1. This section aims to provide some background on the study area. Information on how borders are dealt with from a political and administrative perspective, as well as the work done by the NGO *Alianza por la Solidaridad* [40] in order to make visible and to cover some gaps found in the functioning of the more formal or official governance structures are presented.

---

[3] Territory is therefore understood not as the geopolitically confined area, as defined by State's administrative borders, but as social products with changing borders based on the movement of local actors, the inhabitants of that area.

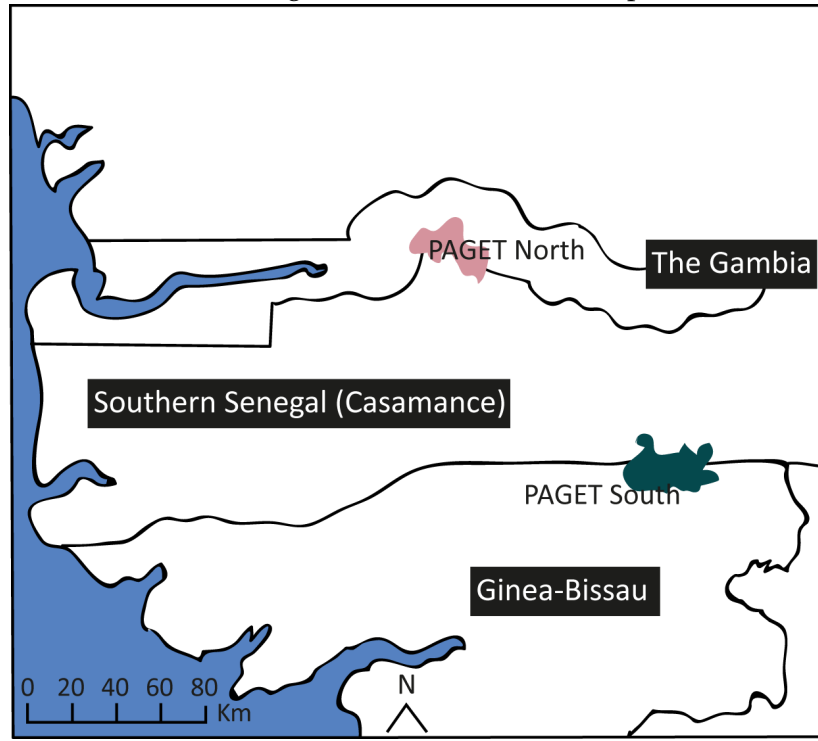

Figure 1: Localizing the Study Area. Source: own elaboration, based on SAGE Agreement.

The study area is one of these African borders where traditional ties survived the European creation of Nation-States and their territorial limits as people living near them remained involved in cooperation networks and shared land and resources management practices (see [41] for more information on pre-border traditional ties and networks). The work of *Alianza* emerged from the observation of a cross-border mobility, built on these traditional networks and interactions, that seemed to represent a shared use of the scarce natural and man-made resources present near the border in both countries. This shared use of resources, despite being the only possibility for many to thrive in the region, far away from national and regional capitals and their development networks, appeared also to be the source of many conflicts due to increasing levels of land degradation and food insecurity. The international NGO's work tried to bridge the gap between the need for a land management system to govern and to arbitrate in the conflicts that emerge out of the common use of resources, and an insufficient, although in progress, cooperation among neighboring local governance structures, hindered by the heterogeneity in their organization and competences.

## 2.1. Conflicts over natural resources: heterogeneous local governance structures, a barrier to cross-border land management and planning

The study area has traditionally lived on agriculture. However, higher population densities in the area increase the pressure over natural resources[4], leading to their degradation and threatening certain socio-economic balances: conflict between farmers and stock-breeders, in addition to preexistent political issues[5], contribute to perpetuate the relative regional instability, in addition to threatening the main source of nutrition for the local population.

Although difficult, cross-border cooperation exists in the area: numerous texts, political initiatives and institutions are strongly geared towards the sub-regional integration of these countries. The considered countries, Senegal, the Gambia and Guinea Bissau, are members of the Economic Community of West African States (CEDEAO). Additionally, Senegal and Guinea Bissau belong to the same monetary zone: West Africa Economic and Monetary Union (UEMOA). The adhesion to such frameworks implies the acceptance of free movement of people and goods. Of special relevance have been the harmonization-exercises to control and regulate transboundary flows of forestry products, which are one of the main sources of conflict in these areas. However, cross-border integrated land planning and management as well as natural resources exploitation remain difficult issues given the heterogeneity in the structures of local administrations.

This heterogeneity can be understood as a result of individual and diverse nation-state post-colonial development paths[6]. This translates, among other results[7], into very divergent decentralization processes. While (sub)regional and local elections already exist in Senegal, the Gambia and Guinea-Bissau have not yet consolidated their decentralization processes: elections are only held at central administrative level. Consequently, individuals living in these border areas are subject to very different forms of governance (more or less decentralized, refer to Figure 2), with different approaches and competencies in the management of their shared resources (infrastructure, social services, natural resources, etc.). Under this framework, shared management and cross-border cooperation become extremely difficult; especially if we take into account the limited resources allocated to these border areas, normally rural, far from dominant urban networks and with very low representation in national agendas.

---

[4] The increase in population in southern Senegal, Casamance region, could be explained by the desertification process taking place in the North of the country and its neighbors Mauritania and Mali. North-south migration flows and households' resettlement in the Casamance could be behind the higher pressure on resources, due to an increased need of food-products and wood for cooking.

[5] The Casamance region, separated from Dakar by the Gambia, has traditionally fallen out of the main development trends of the Senegalese central government. The protraction of the abandonment across decades led to the uprising of local population, from a different ethnic group than those in government, claiming for their rights.

[6] Senegal being an ex-French colony, Guinea-Bissau Portuguese and The Gambia British: metropolis with very different (self)development capacity, that introduced different administrate structures, and decided on divergent political and development trajectories for their colonies.

[7] In this section the focus is on administrative differences, later on (see section 4.1) economic development differences are also considered.

## 2.2. Cross-border, participatory land management - the four phases of the PAGET Methodology

Seeking to address these issues of food insecurity, misuse of natural resources and absence of structured and shared planning and management processes, the Spanish Agency for International Development Cooperation (AECID) conceived and implemented the 2011-2016 SAGE Programme (Food Security and Environmental Governance [51]). This programme had four major components: land planning, food security, environment and gender, and counted with the technical assistance of a local organization from each country: Forum pour un Développement Durable Endogène (FODDE) in Senegal, Agency For The Development of Women & Children (ADWAC) in Gambia and Associação para a Promoção do Desenvolvimento Local (APRODEL) in Guinea Bissau. The overall coordination of the programme was granted to *Alianza*.

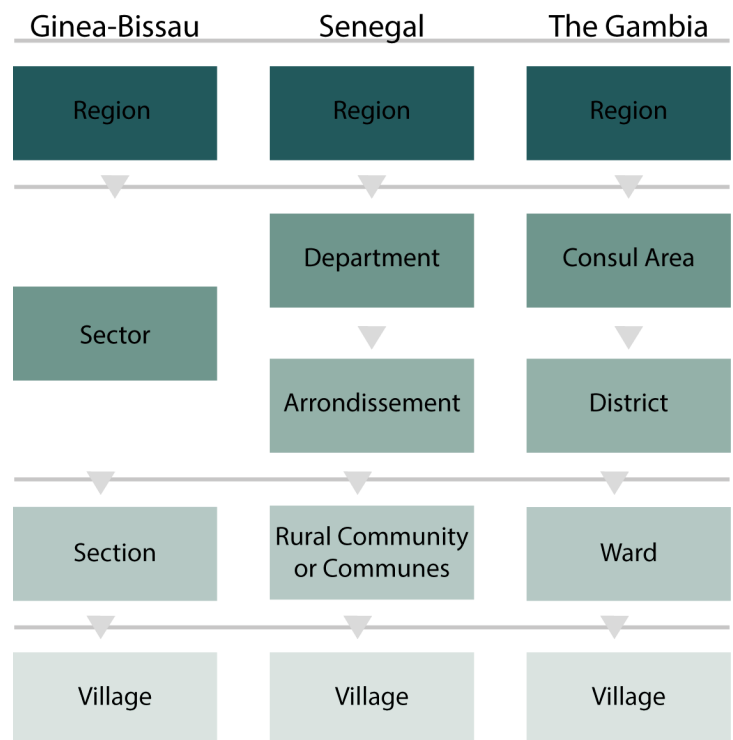

Figure 2: Comparative diagram of the administrative divisions in the three countries. Source: own elaboration based on the SAGE Agreement.

Within the context of SAGE's implementation, a baseline study carried out in the three countries revealed very different territorial morphologies and dynamics, as well divergent land management tools and administrative rules and procedures (see section 2.1). Following these findings, since cross-border collaboration was central to the conception and objectives of SAGE, a forum was organized in Kolda (Senegal) in 2011 with representatives from the technical services, local authorities and civil society organizations of the three countries. The aim of the forum was to agree on and define a common land management approach for cross-border areas. As a result of the forum, the PAGET methodology came to place, and four stages for their implementation identified: preparation, diagnosis, planning and implementation.

In the Preparation Phase two cross-border zones were identified in the sub region: PAGET North (between Senegal and The Gambia) and PAGET South (between Senegal and Guinea Bissau). The PAGET South intervention zone (see Figure 3), object of this study, covers the area between the Wassadou Rural Community (Kolda Region, Senegal) and the Pirada sector (Gabu Region, Guinea Bissau).

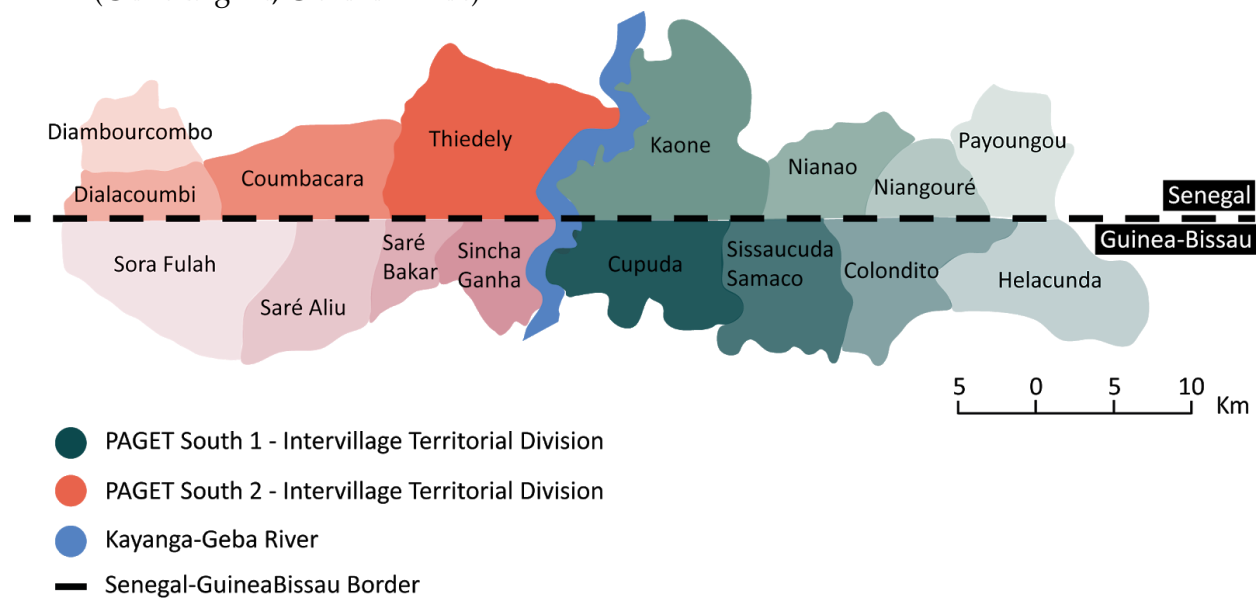

Figure 3. Study Area, PAGET South (see Figure 1) at the border between Senegal and Guinea-Bissau. Source: own elaboration, based on SAGE Agreement.

In parallel, to ensure qualitative representation of the local population, a cross-border land management committee (see Figure 4) was established following the basic principles of the methodology: real local control, responsibility, democratic representation, decentralization, local capacity/resources promotion, gender equity and equality, and consultations and active participation of all intervening actors in the process. Pre-existing structures were reviewed with the support of local authorities and partners, harmonized into a final proposal with committees at four different levels: village, inter-village, land committee level and cross-border PAGET. The connection with formal administrative structures was made through the land committees and the PAGETs, which worked in collaboration with the technical services of State's lowest administration levels, who played a consultative and monitoring role.

During the Diagnosis Phase, a team, led by the PAGET (sub)committees supported by technical experts provided by *Alianza* and the local NGOs (FODDE, ADWAC and APRODEL), carried out a field data collection process in close collaboration with local populations: a number of participatory workshops were conducted in the cross-border, inter-village sub-zones to ensure communities' involvement (refer to section 5.1 for further details). Following the consultation exercise, intermediary reports were shared with local populations to capture their feedback and inputs before the drafting of the final report. Lastly, a final cross-border diagnosis report, and corresponding visualizations, was submitted for the validation of the PAGET Committee in presence of technical services and local authorities. Overall, this phase helped to develop a shared vision of development issues, constraints, potentialities and main trends in the management of cross-border spaces.

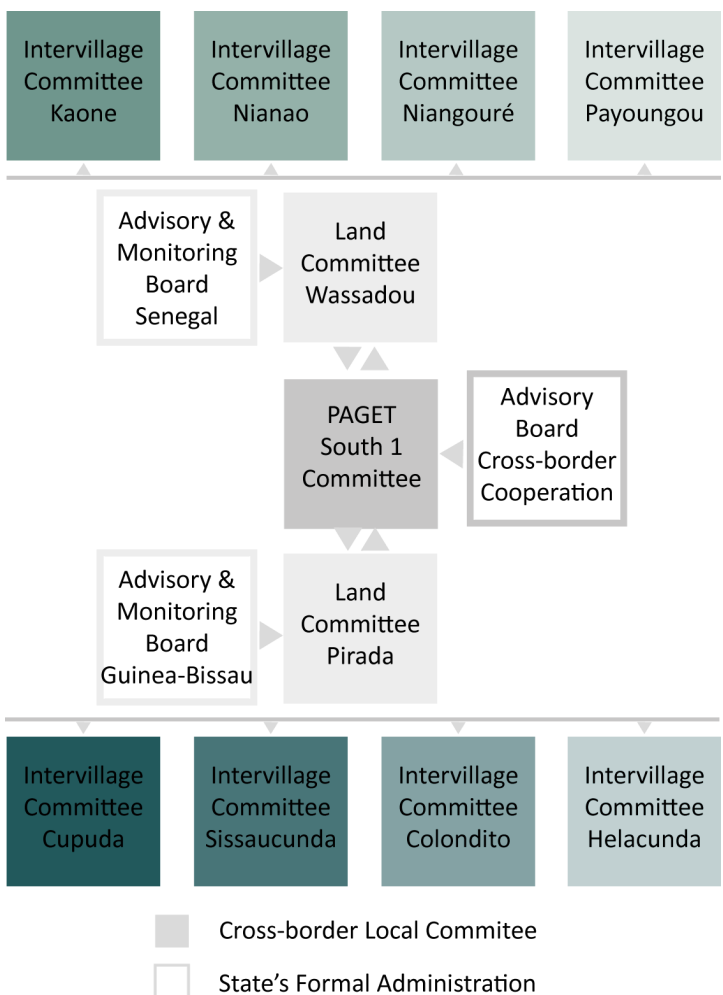

Figure 4: Complementary Cross-Border Land Management Structure developed within the SAGE Program. Source: own elaboration based on SAGE Agreement.

Based on the Final Diagnosis, during the Planning Phase, a list of interventions was elaborated and scheduled for implementation within a 5 years horizon. All planned interventions were categorized under six strategic axes: (1) equal access to and control over the land, (2) more efficient, gender-sensitive production systems, (3) natural resources' preservation and restoration, (4) sustainable management and exploitation of transboundary forestry resources, (5) free movement of goods and people in cross-border areas, (6) strengthen cooperation dynamics through an harmonized transborder local development agenda.

The SAGE program ended with the elaboration of the plans PAGET and the corresponding transborder maps, their Implementation Phase remained out of the financial capacity of the initiative. To help local communities and cross-border committees in the materialization of these plans, *Alianza* supported political efforts to pressure local and national State's authorities to integrate the PAGET plans into their development agendas. A positive result of this activism was the creation of the Programme d'Urgence de Modernisation des Axes et Territoires frontaliers (PUMA) [46] in Senegal.

In parallel, *Alianza* also worked to attract international investors that could contribute in the development of the needed infrastructure that had to be developed in order to achieve the agreed strategies in the Planning Phase, in particular axis one, four and five. It was within this context that *Alianza* signed an agreement with the *Civil Engineering School of the Polytechnic University of Madrid* (UPM) for the design and implementation of the infrastructure projects that had been identified as priorities. During the development of one of these projects: a bridge over the Kayanga-Geba River at the border between Senegal and Guinea-Bissau, a question about the real demand for its construction, as well as about its optimal location and design parameters, in a context of very limited resources and tight budget, was raised. To prove the necessity and suitability of the bridge for local development, a transborder trip generation and distribution model was conceived for the PAGET South area. The development of such a model was the main motivation behind this research.

## 3. Purpose of the study: guiding research questions and aimed contributions

As indicated in the previous section, the study emerged as a means to defend the construction of the bridge as a necessity for local development. Hence, the research question behind this paper's work was: *What is the impact of the hypothetical construction of the bridge on the Kayanga-Geba river on South PAGET's cross-border mobility?*

Nonetheless, in order to contribute to the legacy of the work done under the SAGE Programme: to support the acceptance and incorporation of the PAGET plans into local and regional development agendas and to help in the development efforts to improve the well-being of local populations, three detailed research goals were considered:
- To understand to what extent the present case study can contribute to the existing literature concerning cross-border (rural) mobility modeling, visualization and analysis.
- To initiate a reflection around which and how data (such as that gathered through the different phases of the PAGET Process), in particular that related to mobility, could be visually represented to simplify the spatial analysis of the study area, contributing to a better informed, participatory, need-based governance of cross-border areas.
- To grow the number of precedents of mobility models and visualizations conducted in development contexts, in an attempt to shed some clarity on how alternative trajectories can be integrated into pre-existing, northern-developed models.

To better understand the last bullet-point, one last remark should be raised in relation to the negative consequences that come from a decontextualized application of technology and methodologies. Porter [45] and Levy [35, 36] argue that most urban transport planning and development schemes, as well as the technology used to perform those, are based on hypotheses coming from the Global North. Hypothesis built on mobility patterns that often do not match the reality within African cities and that, furthermore, cover multiple social inequalities that have already led to unsustainable forms in Global North countries, challenging their suitability as development drivers elsewhere. Following the authors' reflection, coming out of the urban context, with this paper we also aim to contribute to a better understanding of the processes that need to be in place in order to ensure that the application of foreign technology and methodologies does contribute to sustainable local development instead of to the introduction or reinforcement of inequalities.

## 4. Related works

**4.1 Cross border mobility[8]**

Most of the research on transborder mobility focuses on trips motivated by the need to access healthcare systems. One reason behind this trend could be derived from Tapia Ladino et al.'s [54] data. In their study concerning the motivations behind cross-border mobility along the chilean-peruvian border the authors found that 40 percent of the trips were due to health tourism, the greatest among the enquested options.

Among the bulk of research that deals with health-motivated transborder mobility, *border asymmetries* in the access to services seems to emerge as one of the predominant reasons, a possible interpretation of Massey's [37] power-geometries of relations as the force behind territorial development. So it shows the research carried on the Lao-Thailand [6, 7] and Haiti-Dominican Republic [38] borders, where is the greater quality of the health systems in Thailand and the Dominican Republic, countries with a higher level of development than their neighbours (Thailand's 0.765 versus Lao's 0.604 and Dominican's 0.745 versus Haiti's 0.503, according to 2019 UNDP's Human Development Index [24]), what triggers the movement of patients from the border regions of Laos and Haiti. The other reasons mentioned are the centripetal forces of historical ties and proximity, understood as physical accessibility. As a result of these factors, between 7 to 25 percent of the Laotians living in border areas seek health treatment in Thai facilities, while 10 percent (35 percent in primary care centers) of the patients treated in borderland Dominican health infrastructures are Haitian foreigners.

In the case of the area under study in this paper, Senegal shows greater levels of development than neighboring Guinea Bissau: 0.514 versus 0.461 according to 2019 UNDP's Human Development Index. Therefore, linking the notion of *border asymmetries* to our study area, it would be logical to assume a South-North mobility trend due to the potentially better services available in the Senegalese side of the border. An hypothesis to be validated through the mobility study.

---

[8] In this section we acknowledge the difficulties and limitations that emerge from the comparison of very different geographies. However, we enter the discussion on related works from Robinsons's [48] perspective, who sees in the introduction of comparative tactics a useful tool for the emergence of conceptual innovation, although minding that concepts are always open to revision.

Cross-border mobility has also been studied from a livelihoods perspective. Raeymaekers [47] study of livelihoods across the Republic of the Congo-Uganda border brings interesting insights on the connection between transborder mobility and governance. In his work, Raeymaekers reflects on the role that cross-border everyday engagement has on state-making, on the construction of power on peripheral areas. His argument builds on the hypothesis that trade cross-border practices, presumably informal, have the potential to transform the state through constant outwitting: "push(ing) the states to mediate, rather than impose, political authority" (pp. 9). With this analysis, Raeymaekers wants to transcend dominant discourses on borderland, getting away from liberal interpretations of these areas as a "distortion of rational economic rules" or Marxist assumptions on border territories as spaces of resistance or on the state of exception. In an attempt to establish a more realistic perspective, Raeymaekers sees in "borderland a crucial pillar in the equilibrium forces between formal and informal, state and non state actors and regulations in this territorial periphery" (pp.6). Raeymaekers' perspective on borderlands will be taken as a framework for this paper, especially when discerning how mobility analysis and visualization can support local (informal) cross-border governance systems, as the land management structure developed within SAGE, to mediate its legitimacy, against and with national (formal) structures, as a land management authority able to effectively arbitrate the conflicts emerging from a shared transnational use of resources and services.

**4.2 Mobility modeling**

Modelling mobility in rural areas is not easy, mostly due to the difficulties in acquiring the appropriate data, e.g., data on travel characteristics in rural areas are usually missing, while in some cases, e.g., in developing African countries, there is even missing information about available road infrastructure [33]. Therefore, difficulties in the proper modelling comes from lack of data for both, the demand and the supply side of transportation modelling. Therefore, designing new infrastructure is a challenging task. Traffic modelling and simulations have an irreplaceable role in transport infrastructure planning [17] but data scarcity requires introduction of simplifying assumptions allowing for simpler and tailored solutions.

In the existing literature, the rural road network planning methodology is usually based on the concept of accessibility, defined by scholars usually in terms of the possible use of the services (potential accessibility) and their actual use (realized accessibility) [38]. For example, in Singh [52], a new index of accessibility is designed which evaluates various rural road link options for their efficiency in accessing the missing functions in the unconnected settlement. The accessibility-based approach of rural road planning offers maximum benefit to the unconnected settlement in terms of access to various facilities or the main road network in a coordinated fashion by maintaining an integrated road system. Also, in Briceño-Garmendia [11] accessibility is modeled using a gravity model [13] inspired by the Newton's Universal Law of Gravitation, which assumes that the number of trips between zone A and zone B is proportional to the number of trips originating from A and destined for B, and inversely proportional to the function of distance between A and B. This function is a negative power of a distance between A and B, and the exponent is usually assigned by calibrating the model based on real-world data.

In the case of the presented research, due to the lack of data concerning real routes it was decided to use a gravity model to estimate the number of trips between different zones. The gravity model was selected in this case because it doesn't require data about the real routes as an input and can still give some useful outcomes (the details are presented in Section 5.2).

**4.3 Mobility Visualization**

From a visualization perspective, the understanding of mobility patterns, as a form of flow data, can benefit from the use of cartographic presentation methods. Maps help to visualize and understand spatial data, as they highlight spatial relations and patterns of phenomenon distribution [32]. By visualization, we understand assigning appropriate graphic signs to particular objects. In this context, the visualization can be seen as a graphic recording of spatial information [42].

In Alcorn [2] opinion, the analytical power of mapping techniques creates a possibility for NGOs to improve local decision-making and to provide convincing evidence to outsider audiences and development actors. Maps are a useful tool for (self)governance as sources of quickly readable, already summarized information [2]. Additionally, visualization brings the possibility to explore future scenarios, to analyse changes before they take place, thus useful for communities and authorities to decide among development options [5, 50].

In order to select the most suitable visualization method, Dent et al. [15] distinguishes between thematic and topographic maps as a starting point. On thematic maps we can find information about spatial characteristics of the presented phenomena. Topographic maps show objects like houses, reference grid, relief. As flow data show patterns and trends of the phenomena they can benefit most from the use of thematic visualizations [15]. Choosing an appropriate cartographic presentation method depends on the nature and characteristics of the data that we want to visualize. In terms of nature, there exists relative and absolute data, which can be shown either in continuous or discrete ways. When it comes to characteristics, cartographic presentation methods are also conditioned by the data measurement level (quantitative, qualitative, ordinal) [31a, 31b, 42]. Thematic maps are designed with the use of qualitative and quantitative presentation methods. Qualitative methods are usually used to show location, spatial distribution of the nominal data. Quantitative methods represent numerical data and their spatial aspects [15].

The graphic visualization of aforementioned different aspects of the data sets is obtained through the application of Bertin's [4] six graphic variables: size, hue (colour), brightness, orientation, graininess, shape. Hence, Bertin's variables have a great influence on maps' graphical characteristics. Size is usually linked to the presentation of quantitative or ordinal data. Qualitative characteristics are usually visualized through color, orientation, graininess and shape. Brightness allows data to be differentiated at ordinal measurement level: after proper organization, brightness creates the impression of a change in the intensity of the phenomenon [31a, 31b, 42].

In the literature, various types of cartographic presentation forms used for the visualization of flows can be found. For instance, two colours straight lines [10], or circular migration plots [49]. Meanwhile, Pasławski [42] and Pieniążek et al. [43] recognize line diagrams as a cartographically appropriate form for movement and flow representation. The later authors distinguish between two types of line diagrams: vector and ribbon; while ribbon diagrams take into account topographic objects (e.g., road networks), the second type simplifies the visualization to the use of a line connecting points or areas, without further topographic considerations [43]. Most of the available literature defines both types of diagrams with the term "flow lines" [53] or generally "flow maps" [10].

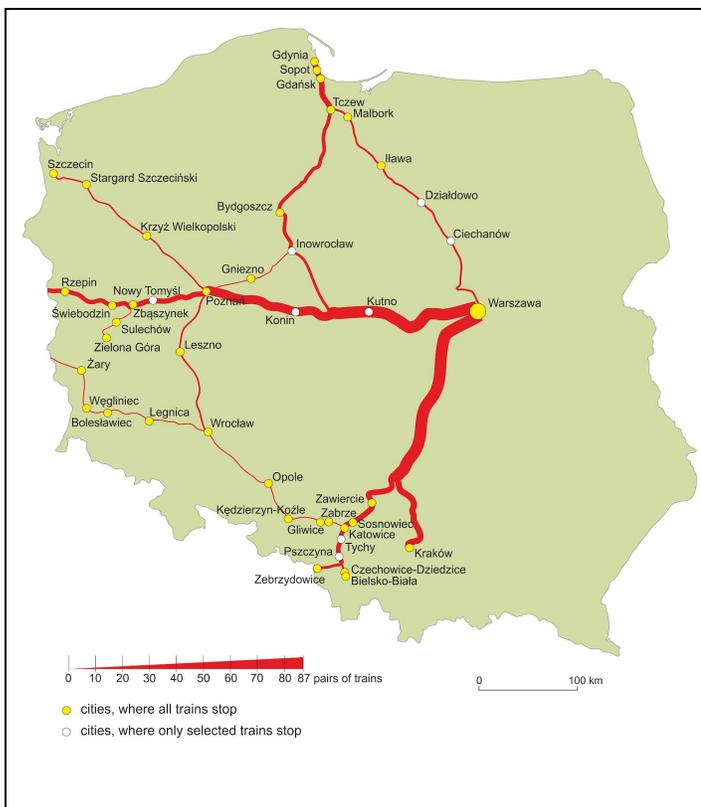
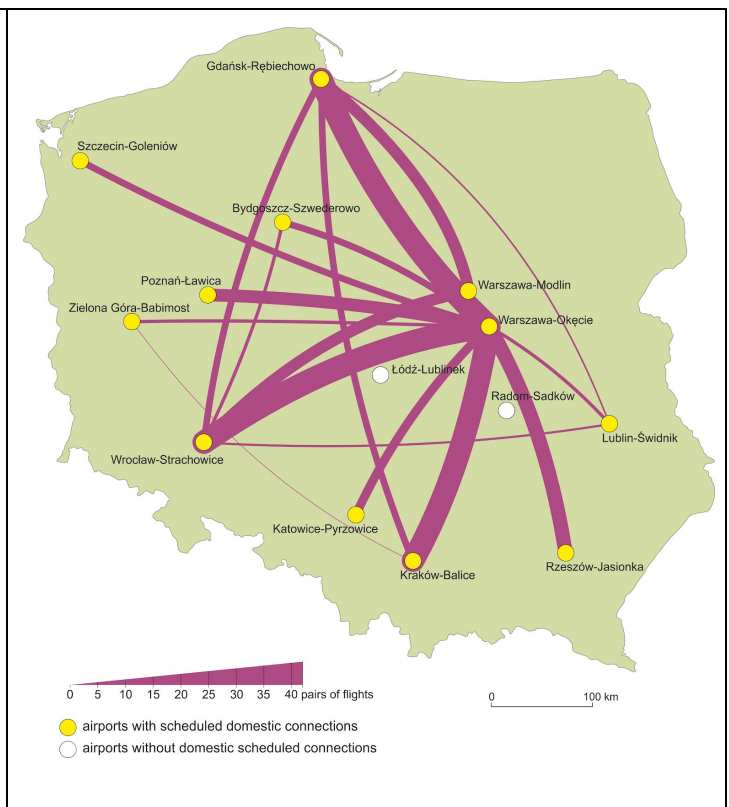

| Figure 5. Ribbon linear diagram map. Source: Pieniążek and Zych [44]. Published with permission. | Figure 6. Vector diagram map. Source: Pieniążek and Zych [44]. Published with permission. |
|---|---|

Boyandin et al. [10] noticed that the most popular form of visualizing movement are flow maps. In these maps, contrary to possible preconceived assumptions, flows usually do not imitate migration paths: flows do not follow the topographic networks (i.e., roads, railways etc), rather focusing on the representation of the start-end poles of the movement (sources and destinations of migrations), showing the general trend of the movement. Additionally, flow maps can also help to visualize the smallest and the largest concentrations of flows connecting different migration poles, allowing to observe differences in the volume of the flows. In order to show the above, flows are represented by lines connecting the start with the end point, e.g. start and end of journey [10]. These connections of points are represented for quantitative data, such as flow intensity, as lines of increasing thickness. For qualitative data, for instance the types of infrastructure or different road categories, the hue may be used [53].

Among the disadvantages of flow mapping methods is the decrease of the quality of data's visibility due to overlapping and crossing of lines [15, 26]. This may cause the problem referred to in the literature as 'visual clutter', connected with too small line spacing while visualizing big data sets [10, 30, 58]. To avoid this issue, smaller flows should be located above larger ones [15]. An additional difficulty in such representation is the tendency of accumulation of flows in large towns and decrease in smaller villages [49]. However, this problem can be partially solved by data aggregation and generalization [9]. For example, a threshold value for visible flows can be proposed to limit the clutter and efficiently present the information on the map [55, 23a, 23b]. These processes, known as 'cartographic generalization', lead to the identification of characteristics and key features of the presented phenomenon [12]. Cartographic generalization is necessary for optimal visualization of information, especially of its spatio-temporal (speed change, intermediate points on the route) and spatial (genesis, destination, route) nature [3].

# 5. Methodology

The motivation for this paper's work was to model, visualise and analyse populations' mobility within the considered area, PAGET South, at the border between Senegal and Guinea-Bissau, in order to better understand the potential impact on mobility that building a bridge in the proposed location, over the Kayanga-Geba river, might bring. To this end, first, mobility models for both cases (with and without the bridge) were developed. Due to low data availability, the decision was made to construct a mobility model based on the so-called gravity model [13], in which the origin-destination matrix (OD matrix), with information about the estimated number of trips between particular areas, is developed based on the estimates of trip productions, attractions and distances between areas. In order to reliably compare the results and impact of the infrastructure decision to build a bridge, we decided to calculate OD matrices for the number of trips within a year. Later, those matrices were visually represented using flow maps designed in open source JFlowMap and commercial ArcMap programs.

## 5.1 Datasets development - an outcome of PAGET's Diagnosis Phase

The datasets used in the mobility study come from the gathered information through the Diagnosis Phase of the PAGET Programme. In particular, the data used was: population and settlement location used as production poles and basic infrastructure (school, hospitals and markets) as attraction poles.

As described in Section 3, in order to avoid the negative consequences that derive from a decontextualized application of technology and methodology, we believe it is important to understand how this data was collected and the role that the PAGET committees and local population had in this process.

The collection and validation of the data was accomplished through participatory consultation and mapping led by *Alianza* in coordination with the local NGOs: APRODEL,

FODDE and ADWAC. A five-steps methodology was followed: (1) collection of primary data through participatory workshops, (2) collection of secondary data, statistics and existing plans and programmes, (3) elaboration of the diagnostic cartography and development of a sub-regional GIS infrastructure, (4) systematization, analysis and processing of data, and (5) drafting and validation of the diagnostic report. In the next paragraphs we describe steps 1 and 3, through which the data used in this study was gathered and georeferenced.

During **step 1**, a number of participatory workshops were delivered at the different PAGET's (sub)committees, starting from the lowest one: villages, to then move up to inter-villages, land and PAGET committees in the end. Groups were organised according to professional expertises and, in order to ensure a gender-sensitive analysis, information was always disaggregated by sex and specific women focus groups were created. Five different diagnostic tool were used as guiding parameters of the analysis:

1. Analysis of socio-cultural spatial trends. Through a participatory mapping exercise participants were asked to place or draw historical cross-border ties and border crossing points, basic infrastructures (hospital, eschools), commercial infrastructures (markets), civil infrastructure and natural resources. As a result of this activity the facilitators gained a good understanding on the different land uses as well as the location of the different man-made and natural resources.
2. Analysis of socioeconomic trends and gender inequalities. The tools used in this case were: income and expenditure matrix, seasonal calendar, table of resources' access and control and daily routine clock.
3. Analysis of actors and dynamics of cross-border flows: Venn diagram and cross-border flow diagram.
4. Thematic workshops on environment and food safety
5. Classification and prioritisation of problems through a SWOT and problem tree analysis.

In the **3rd step**, an open source (Geographical Information System (GIS) was put in place, synthesis of all data collected through the two previous steps. The building-up process for the GIS system was the following:
- All hand-drawing maps done with the communities were interpreted and mapped to scale. The final result was printed and shared with the participants of the workshops for their feedback.
- Then, a database was created with the geolocation of the fundamental elements of the cross-border area, including: main forests, valleys, agricultural areas, basic infrastructures (schools, hospitals and markets) and road networks. This database was then used for the creation of a number of maps that configured the diagnostic and planning cartography of the program.

Out of the PAGET Programme, for the purposes of this study, the data gathered by the Programme was summarized in an Excel spreadsheet [14], which was the main dataset used in the study. As mentioned above, only settlements' locations and population density and infrastructure (hospital, schools and markets) were included; social ties, despite their already mentioned influence on mobility, were left out due to the difficulties to translate them to quantitative terms, able to be computed into the model in the form of production/attraction poles. Similarly, not enough gender disaggregated data was available to analyze gender influences on regional mobility. Further reflection on these limitations can be found in Sections 6.3 and 7.1. Georeferenced information about the road network was not made available for use of the authors.

## 5.2 Mobility modelling

In transportation engineering, one of the most popular models for mobility prediction is the Four Step Model [39]. It assumes that the study area is divided into zones and its two first steps are trip generation (which determines the frequency of origins or destinations of trips in each zone by trip purpose, as a function of land usage and household demographics, and other socio-economic factors) and trip distribution (matches origins with destinations). The other 2 steps are mode choice and route assignment. However, while conducting our research, we didn't have access to information about possible modes of transport nor the exact routes. Therefore, we decided to model mobility using only the first 2 steps which let us build the origin-destination matrices for the study area. As mentioned in Section 4.2, a good way to go in such cases is to apply the so-called gravity model [13]. However, in our case, we didn't have access to any such real-world mobility data which could be used to verify the model, so we decided to apply an exponent value based on analyzing similar case studies in the literature [25]. It turned out that the proper value is usually in a range [0.5, 3] and since we didn't have any additional evidence on which value may be the best, we decided to use the value 2 just to simplify the computations. This gave us the following formula:

$$Q_{ij} = k * P_i * A_j / W_{ij}^2, \qquad (1)$$

where:

- $Q_{ij}$ is the number of trips between zones "i" and "j"
- $P_i$ is the number of trips produced by zone "i"
- $A_j$ is the number of trips attracted by zone "j"
- $W_{ij}$ is the "inter-zone impedance", the difficulty/resistance to the trips between zones "i" and "j". Usually, the "trip time" or "trip distance" is used as a reference value. Since we didn't have access to information about the exact routes and road network structure, we decided to apply a distance in a straight line between zones "i" and "j".
- "i" and "j" are two variables to be estimated through model calibration based on annual statistics
- "k" - proportionality constant

However, k is commonly disregarded and we are left with the following formula:

$$Q_{ij} = P_i * A_j * F_{ij} / \sum_x (A_x * F_{ix}), \qquad (2)$$

where $F_{ij}=1/W_{ij}^2$. In order to calculate $Q_{ij}$, we estimated the productions and attractions due to 3 types of infrastructures: schools, hospitals and markets. In the case of schools, we considered kindergartens, primary schools and secondary schools. We estimated the number of trips due to each of the infrastructure types separately, so the total number of trips was just an aggregation of OD matrices for different infrastructure types.

**5.3 Adaptation of the gravity model**

In order to build an OD matrix using the gravity model for the investigated case, the following process was used to determine each of the parameters.

Firstly, the population and infrastructure data (hospitals, schools and markets) of the PAGET South were taken from the information provided by *Alianza* (see Section 5.1). Later, we had to define the zones considered in the OD matrix and their centers. We decided to consider each town or village of our dataset as a separate zone and a town or village center was considered as a center of a zone.

As starting data to define trips, the populations of each of the zones were considered, and the estimated percentage travelling to each of the different infrastructures was calculated. We considered 3 types of infrastructures: schools, hospitals and markets. For each of them we estimated the number of yearly trips from each of the villages. Therefore, we were able to estimate the production part of each village.

For schools, the percentage of Senegal's population pyramid was crossed with the French educational system (model followed in Senegal) to estimate the number of children of school age. In this way, Table 1 was obtained:

*Table 1. Percentage of the population in the school age. Source: Senegal Population Pyramid, World Population Prospects, United Nations, Department of Economic and Social Affairs.*

| French Educational System | % of total population | |
| --- | --- | --- |
| | Male | Female |
| Nursery school (age 1) | 8.7 | 8.5 |
| Kindergarten (ages 3-5) | | |
| Primary school (ages 6-10) | 7.3 | 7.1 |
| Secondary school (ages 11-4) | 11,5 | 11.2 |
| High school (ages 15-17) | | |

We assumed that all children in a given age have to travel to kindergarten or schools, so based on Table 1, the daily number of trips from each village for the purpose of education during a school day can be assumed as:

$P_{Kindergarten,i}$ = 0,085 * female population in zone "i" + 0,087 * male population in zone "i"

$P_{Primary,i}$ = 0,071 * female population in zone "i" + 0,073 * male population in zone "i"

$P_{Secondary,i}$ = 0,112 * female population in zone "i" + 0,115 * male population in zone "i",

where $P_{Kindergarten,i}$, $P_{Primary,i}$, $P_{Secondary,i}$ are daily productions (the number of daily trips during a school day) for zone "i" due to kindergarten, primary schools and secondary schools, respectively. In order to calculate the yearly number of trips, we multiplied these numbers by the approximated number of school days in the considered area - 200 days.

In the case of hospitals, the Figure 7 (from the World Health Organization (WHO) [57]) was used to determine the most frequent causes of use of health services in Senegal:

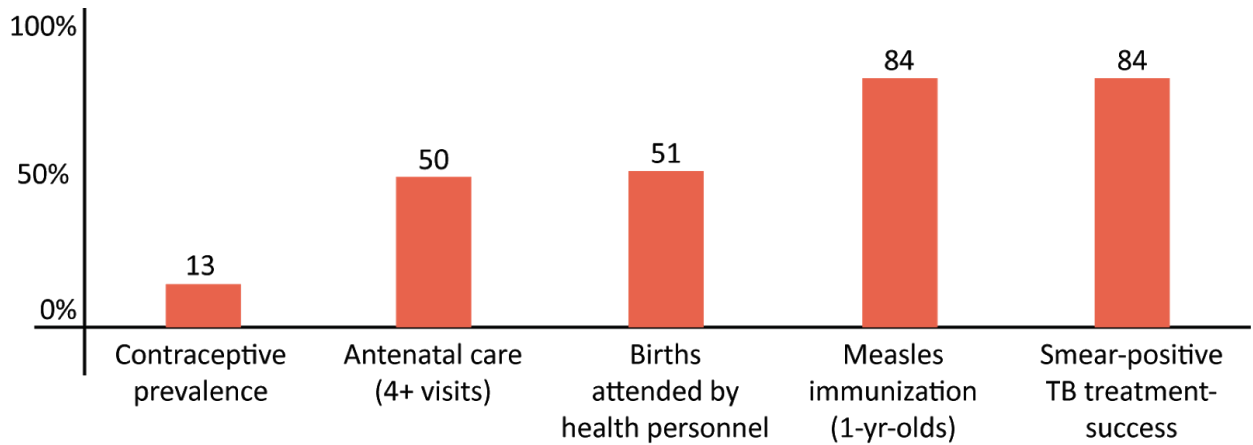

*Figure 7: Utilization of health facilities in Senegal, 2007. Data Source: World Health Organization (WHO) [57].*

Regarding the percentage of the population "affected" by each of the conditions that appear in the Figure 7, the Senegal file provided by WHO on its website has also been used [57]. The numbers are:

- Percentage of the population affected (yearly) by tuberculosis (TB): 0.14% (2015)
- Yearly birth rate (births as a percentage of the population): 3.78% (2013). We assume that this value coincides with the number of annual pregnancies.
- Infant mortality rate (less than a year): 4.39%

Therefore, the yearly production of each zone "i" in relation to health infrastructure can be assumed as:

$P_{Hospital,i}$ = 0,0378*(4*0,5+0,51) * zone's population + 0,84*0,172 * zone's population + 0,0014*0,84 * zone'spPopulation

The first component, 0,0378*(4*0,5+0,51), corresponds to the number of yearly hospital visits due to the pregnancy (as only 51% percent of births are attended by skilled health professionals, we assumed that 50% of women attend hospital during the pregnancy and the number of visits is equal to 4).

The second component, 0,84*0,172, corresponds to the estimated number of visits in the hospital of children being below 5 years old. As we know from Table 1, children in such an age constitute 17.2% of the total population. Only 84% of children attend the measles immunization (in hospitals) and we assumed that the same percentage of children below 5 years old attend hospitals once per year.

The last component, 0,0014*0,84, corresponds to the percentage of the population travelling to hospitals due to tuberculosis.

In the case of markets, it has been considered that trips to markets and other services are made by 25% of the population of each village. We don't have sufficient data to estimate it better, but this value was assumed just to demonstrate the methodology (for practical applications, some additional analysis should be carried out).

To estimate the attraction, we assumed that for each of the considered infrastructure types it is equally divided between all locations with those infrastructures. Therefore, the attraction of the zone "i" due to the infrastructure of type T is equal to:

$A_{T,i}$ = 1/(number of infrastructures of type T within the study area)

While for school facilities the different levels were computed as different types of infrastructure, for hospitals and markets the lack of data obliged to assume all levels as equal and with the same attraction.

Having all the required data, the origin-destination matrices containing information about the estimated number of trips in a year can be computed based on formula (2). According to the proposed methodology, the decision was made to calculate two OD matrices: for scenarios with and without the bridge. In order to perform such calculations, a dedicated module was implemented in the Traffic Simulation Framework software, which is a comprehensive tool for simulating and investigating mobility [22]. The software reads all the required data as an input and applies the formula (2) to perform computations. The results of those computations (e.g., OD matrices for the two investigated cases) are presented in Section 6.1.

**5.4 OD Matrices visualization methods**

As mentioned in Section 4.3, visualization can be an efficient way to explain and analyze flows within the study area. In the last years, some softwares allowing for the automatic elaboration of such visualizations have been developed, e.g. open source programs such as JFlowMap [27] or QGIS' FlowMapper Plugin [19] or commercial programs like ArcMap's *XY to line tool* [1]. However, available flow mapping tools still require manual modifications in order to fulfil cartographic requirements. One of the limitations of designing automated flow visualizations constitutes the limited existence of flow mapping tools within GIS packages [49].

Consequently, in most cases, flow maps are designed manually or the dedicated software is being designed with the use of appropriate programming language.

In the investigated scenario, the designed flow maps aim to represent people's movement between villages in Senegal and Guinea Bissau. A movement previously defined (see previous section) as motivated by the access to schools, hospitals and markets. The visualization of this movement has the overall goal to contribute to the understanding of the importance of building a bridge within the analyzed area.

ArcMap and JFlowMap were the programs selected to prepare the maps. The aim was to show the differences in people's mobility under the two considered hypotheses: with and without the presence of a bridge over the Kayanga-Geba River. Important considerations should be taken into account.

In the visualizations, thematic and background data is considered. Thematic data included OD matrices and the geographic coordinates of the migration poles. The initial processing of the thematic data, both in ArcMap and JFlowMap, included the following stages: (1) selecting the cities to be represented from all included in the OD matrices (two datasets on flows for the two hypotheses into question: with and without the bridge); (2) converting the coordinates from DM (degrees and minutes) to WGS84 coordinate system (3) importing the data into the programs. The third step, namely the data import to ArcMap and JFlowMap follows different approaches. Data for JFlowMap was prepared according to the description available on the Github website [28]. It was two csv files containing information about nodes (code, city name, latitude, longitude) and flow's intensity (code of origin, destination, flow intensity). The codes were letter abbreviations invented by the author of the maps, resulting from the names of places, written in capital letters. For example: SA is Sare Aliu, SAREBA means Sare Bakar. Each village received an individual code. These two csv files are put together in a config file in the jfmv format. Shapefile with a world map was taken from a dataset attached to the program. But due to the scale, the visualizations only show a thin line that forms the border between Senegal and Guinea-Bissau. Thematic data was imported to ArcMap as csv tables (coordinates of the cities and information about flow) and shown using *XY to line* [1]. In relation to the background data, in ArcMap this consisted of the inclusion of shapefiles: the administrative boundaries [8], road and river networks [20]; and csv: distribution of infrastructure (school, hospitals and markets), and population distribution. Background csv files needed to be shown using *XY to point*. Then, for all background data, symbolisation was added. Introducing information about population and infrastructure helps to better understand movement's motivations and its volume. Due to the road network inconsistencies and lacks, these data could not be used for modelling, however, they can support the visualization. We did not find the possibility to include the background information to JFlowMap.

# 6 Results

## 6.1 Computation of the OD Matrices

As mentioned in Section 4.1, the goal was to calculate the OD matrices for two cases: without and with the bridge. In the first case, it was assumed that all the travels are only within the two investigated areas, i.e., without any transport between these areas. Therefore, two OD matrices were obtained: for PAGET SUD 1 (Table 2) and for PAGET SUD 2 (Table 3) . In the latter case, it was assumed that travels between these 2 areas are also possible and the results are presented in Tables 4,5,6 and 7. In the aforementioned tables, the following notation is used: A - Kahone, B - Sare Dembra Asset, C - Nianao, D - Pring Maounde, E - Niangouré, F - Kolondito Fouta, G - Payoungu, H - Cupuda, I - Demoussor Nunco, J - Sissaucunda Samanco, K - Soncocunda, L - Pirada, M - Bajocunda, N - Wasadou, O - Pakour, P - Thidelly, Q - Kandangha Tobo, R - Coumbacara, S - Dialacoumbi Peul, T - Diambour Kombo, U - Sintcha Ganha, V - Sare Bakar, W - Demba Seidi, X - Sare Aliu, Y - Sora Fula, Z - Sare Waly.

Table 2: Origin-destination matrix for PAGET SUD 1.

| O\D | A | B | C | D | E | F | G | H | I | J | K | L | M | N | O |
|---|---|---|---|---|---|---|---|---|---|---|---|---|---|---|---|
| A | 17555 | 0 | 38996 | 14530 | 0 | 0 | 0 | 0 | 0 | 0 | 0 | 0 | 0 | 5196 | 0 |
| B | 26 | 8343 | 19393 | 5939 | 1 | 0 | 1 | 0 | 2 | 0 | 0 | 4 | 0 | 2751 | 0 |
| C | 3199 | 2669 | 128351 | 5147 | 4361 | 14903 | 2965 | 752 | 0 | 1941 | 2142 | 0 | 0 | 0 | 0 |
| D | 11 | 0 | 25045 | 27921 | 4 | 0 | 3 | 0 | 4 | 0 | 0 | 25 | 0 | 2589 | 0 |
| E | 0 | 0 | 55092 | 5800 | 19449 | 0 | 0 | 0 | 0 | 0 | 0 | 0 | 0 | 4166 | 0 |
| F | 0 | 0 | 14068 | 1309 | 14 | 4818 | 2 | 0 | 0 | 0 | 0 | 6 | 0 | 835 | 0 |
| G | 0 | 0 | 61628 | 8316 | 0 | 0 | 22863 | 0 | 0 | 0 | 0 | 0 | 0 | 6536 | 0 |
| H | 6 | 0 | 17377 | 6697 | 1 | 0 | 1 | 7752 | 23 | 0 | 0 | 6 | 0 | 2012 | 0 |
| I | 324 | 81 | 2627 | 1043 | 22 | 55 | 29 | 417 | 6 | 84 | 56 | 0 | 0 | 322 | 0 |
| J | 3 | 0 | 14340 | 6066 | 2 | 0 | 2 | 0 | 2 | 6482 | 0 | 19 | 0 | 1408 | 0 |
| K | 3 | 0 | 34440 | 12046 | 4 | 0 | 4 | 0 | 2 | 0 | 14716 | 59 | 0 | 3037 | 0 |
| L | 2343 | 1373 | 145444 | 35309 | 5156 | 21030 | 2716 | 1107 | 0 | 5323 | 10998 | 317 | 0 | 9296 | 0 |
| M | 2231 | 990 | 25619 | 7805 | 684 | 1439 | 1185 | 1266 | 11 | 965 | 890 | 12 | 0 | 3123 | 0 |
| N | 0 | 0 | 0 | 0 | 0 | 0 | 0 | 0 | 0 | 0 | 0 | 0 | 0 | 0 | 0 |

| O\D | | | | | | | | | | | | | | | | |
|---|---|---|---|---|---|---|---|---|---|---|---|---|---|---|---|---|
| O | 0 | 0 | 0 | 0 | 0 | 0 | 0 | 0 | 0 | 0 | 0 | 0 | 0 | 0 | 0 | |

Table 3: Origin-destination matrix for PAGET SUD 2.

| O\D | P | Q | R | S | T | U | V | W | X | Y | Z |
|---|---|---|---|---|---|---|---|---|---|---|---|
| P | 41387 | 0 | 64868 | 0 | 0 | 0 | 0 | 0 | 0 | 0 | 0 |
| Q | 2162 | 25 | 7650 | 0 | 0 | 0 | 1109 | 0 | 0 | 257 | 160 |
| R | 0 | 0 | 136453 | 0 | 0 | 0 | 0 | 0 | 0 | 0 | 0 |
| S | 1233 | 2 | 31210 | 0 | 73 | 0 | 1181 | 0 | 0 | 4293 | 9732 |
| T | 901 | 0 | 15769 | 0 | 51 | 0 | 764 | 0 | 0 | 2542 | 3099 |
| U | 465 | 3 | 2995 | 0 | 0 | 0 | 826 | 0 | 0 | 111 | 64 |
| V | 0 | 0 | 15328 | 0 | 0 | 0 | 9779 | 0 | 0 | 0 | 0 |
| W | 434 | 0 | 9084 | 0 | 0 | 0 | 1957 | 0 | 0 | 336 | 117 |
| X | 358 | 0 | 8410 | 0 | 1 | 0 | 839 | 0 | 0 | 1392 | 306 |
| Y | 0 | 0 | 11185 | 0 | 0 | 0 | 0 | 0 | 0 | 7136 | 0 |
| Z | 1 | 1 | 10602 | 0 | 8 | 0 | 2 | 0 | 0 | 20 | 6723 |

Table 4: Origin-destination matrix between villages A-M for the scenario with a bridge.

| O\D | A | B | C | D | E | F | G | H | I | J | K | L | M |
|---|---|---|---|---|---|---|---|---|---|---|---|---|---|
| A | 17555 | 0 | 34807 | 14530 | 0 | 0 | 0 | 0 | 0 | 0 | 0 | 0 | 0 |
| B | 24 | 8343 | 18369 | 5939 | 1 | 0 | 1 | 0 | 2 | 0 | 0 | 4 | 0 |
| C | 3141 | 2621 | 128351 | 5053 | 4282 | 14633 | 2912 | 738 | 0 | 1905 | 2103 | 0 | 0 |
| D | 10 | 0 | 24308 | 27921 | 4 | 0 | 3 | 0 | 3 | 0 | 0 | 24 | 0 |
| E | 0 | 0 | 54374 | 5800 | 19449 | 0 | 0 | 0 | 0 | 0 | 0 | 0 | 0 |
| F | 0 | 0 | 13952 | 1309 | 14 | 4818 | 1 | 0 | 0 | 0 | 0 | 6 | 0 |
| G | 0 | 0 | 59715 | 8316 | 0 | 0 | 22863 | 0 | 0 | 0 | 0 | 0 | 0 |
| H | 5 | 0 | 13687 | 6697 | 1 | 0 | 1 | 7752 | 19 | 0 | 0 | 5 | 0 |
| I | 282 | 70 | 1816 | 1032 | 19 | 48 | 25 | 362 | 6 | 73 | 49 | 0 | 0 |
| J | 3 | 0 | 13430 | 6066 | 2 | 0 | 2 | 0 | 2 | 6482 | 0 | 18 | 0 |

| | | | | | | | | | | | | |
|---|---|---|---|---|---|---|---|---|---|---|---|---|
| K | 3 | 0 | 32817 | 12046 | 4 | 0 | 3 | 0 | 2 | 0 | 14716 | 56 | 0 |
| L | 2310 | 1353 | 143099 | 35239 | 5083 | 20731 | 2678 | 1091 | 0 | 5247 | 10841 | 317 | 0 |
| M | 589 | 261 | 11992 | 7107 | 180 | 380 | 312 | 334 | 2 | 255 | 235 | 2 | 0 |

Table 5: Origin-destination matrix from villages A-M to N-Z for the scenario with a bridge.

| O\D | N | O | P | Q | R | S | T | U | V | W | X | Y | Z |
|---|---|---|---|---|---|---|---|---|---|---|---|---|---|
| A | 5196 | 0 | 0 | 0 | 4189 | 0 | 0 | 0 | 0 | 0 | 0 | 0 | 0 |
| B | 2750 | 0 | 0 | 1 | 1023 | 0 | 0 | 0 | 0 | 0 | 0 | 0 | 0 |
| C | 0 | 0 | 196 | 0 | 141 | 0 | 0 | 0 | 192 | 0 | 0 | 92 | 69 |
| D | 2589 | 0 | 0 | 1 | 735 | 0 | 0 | 0 | 0 | 0 | 0 | 0 | 0 |
| E | 4166 | 0 | 0 | 0 | 718 | 0 | 0 | 0 | 0 | 0 | 0 | 0 | 0 |
| F | 835 | 0 | 0 | 0 | 116 | 0 | 0 | 0 | 0 | 0 | 0 | 0 | 0 |
| G | 6536 | 0 | 0 | 0 | 1912 | 0 | 0 | 0 | 0 | 0 | 0 | 0 | 0 |
| H | 2012 | 0 | 1 | 2 | 3690 | 0 | 0 | 0 | 1 | 0 | 0 | 0 | 0 |
| I | 322 | 0 | 44 | 0 | 838 | 0 | 0 | 0 | 56 | 0 | 0 | 14 | 9 |
| J | 1408 | 0 | 0 | 0 | 909 | 0 | 0 | 0 | 0 | 0 | 0 | 0 | 0 |
| K | 3037 | 0 | 0 | 1 | 1622 | 0 | 0 | 0 | 0 | 0 | 0 | 0 | 0 |
| L | 9296 | 0 | 208 | 0 | 2503 | 0 | 0 | 0 | 224 | 0 | 0 | 108 | 81 |
| M | 3118 | 0 | 658 | 3 | 14571 | 0 | 14 | 0 | 719 | 0 | 0 | 1834 | 3649 |

Table 6: Origin-destination matrix from villages N-Z to A-M for the scenario with a bridge.

| O\D | A | B | C | D | E | F | G | H | I | J | K | L | M |
|---|---|---|---|---|---|---|---|---|---|---|---|---|---|
| N | 0 | 0 | 0 | 0 | 0 | 0 | 0 | 0 | 0 | 0 | 0 | 0 | 0 |
| O | 0 | 0 | 0 | 0 | 0 | 0 | 0 | 0 | 0 | 0 | 0 | 0 | 0 |
| P | 0 | 0 | 35575 | 28673 | 0 | 0 | 0 | 0 | 0 | 0 | 0 | 0 | 0 |
| Q | 981 | 309 | 5402 | 3400 | 79 | 181 | 117 | 310 | 0 | 162 | 128 | 0 | 0 |
| R | 0 | 0 | 42461 | 36419 | 0 | 0 | 0 | 0 | 0 | 0 | 0 | 0 | 0 |
| S | 645 | 280 | 18558 | 12365 | 169 | 362 | 289 | 346 | 1 | 250 | 225 | 1 | 0 |
| T | 362 | 155 | 8504 | 6041 | 87 | 189 | 148 | 189 | 0 | 132 | 118 | 0 | 0 |
| U | 269 | 83 | 2016 | 1393 | 29 | 67 | 40 | 232 | 3 | 78 | 58 | 0 | 0 |
| V | 0 | 0 | 8346 | 6946 | 0 | 0 | 0 | 0 | 0 | 0 | 0 | 0 | 0 |
| W | 96 | 36 | 3788 | 3264 | 17 | 39 | 27 | 63 | 0 | 34 | 28 | 0 | 0 |
| X | 110 | 43 | 3683 | 3062 | 22 | 50 | 36 | 72 | 0 | 42 | 36 | 0 | 0 |
| Y | 0 | 0 | 6336 | 4799 | 0 | 0 | 0 | 0 | 0 | 0 | 0 | 0 | 0 |
| Z | 0 | 0 | 6513 | 4471 | 0 | 0 | 0 | 0 | 0 | 0 | 0 | 0 | 0 |

Table 7: Origin-destination matrix between villages N-Z for the scenario with a bridge.

| O\D | N | O | P | Q | R | S | T | U | V | W | X | Y | Z |
|---|---|---|---|---|---|---|---|---|---|---|---|---|---|
| N | 0 | 0 | 0 | 0 | 0 | 0 | 0 | 0 | 0 | 0 | 0 | 0 | 0 |
| O | 0 | 0 | 0 | 0 | 0 | 0 | 0 | 0 | 0 | 0 | 0 | 0 | 0 |
| P | 12293 | 0 | 41387 | 0 | 61903 | 0 | 0 | 0 | 0 | 0 | 0 | 0 | 0 |
| Q | 1324 | 0 | 947 | 25 | 5197 | 0 | 0 | 0 | 485 | 0 | 0 | 112 | 70 |
| R | 15606 | 1 | 0 | 0 | 136453 | 0 | 0 | 0 | 0 | 0 | 0 | 0 | 0 |
| S | 5489 | 1 | 1045 | 2 | 27785 | 0 | 68 | 0 | 1001 | 0 | 0 | 3639 | 8246 |
| T | 2669 | 0 | 748 | 0 | 14426 | 0 | 51 | 0 | 634 | 0 | 0 | 2110 | 2572 |
| U | 494 | 0 | 213 | 1 | 2117 | 0 | 0 | 0 | 378 | 0 | 0 | 50 | 29 |
| V | 2821 | 0 | 0 | 0 | 14599 | 0 | 0 | 0 | 9779 | 0 | 0 | 0 | 0 |
| W | 1348 | 0 | 399 | 0 | 8831 | 0 | 0 | 0 | 1797 | 0 | 0 | 309 | 107 |
| X | 1278 | 0 | 321 | 0 | 8103 | 0 | 1 | 0 | 752 | 0 | 0 | 1248 | 274 |
| Y | 2078 | 0 | 0 | 0 | 10658 | 0 | 0 | 0 | 0 | 0 | 0 | 7136 | 0 |
| Z | 1975 | 0 | 1 | 0 | 9663 | 0 | 8 | 0 | 2 | 0 | 0 | 19 | 6723 |

## 6.2 Visualizations of the results

Two maps in ArcMap and two visualizations in JFlowMap, concerning flows with and without the bridge, were designed.

For Figure 8, prepared with the use of ArcMap, background content was prepared using a signature method. The different types of infrastructures linked to the visualized flows (education, health and commerce) were differentiated at ordinal level using changes in hue and colour: objects from the same group (e.g., educational infrastructure: kindergarten, primary schools) were assigned the same colour (green) with different hue. Population, as quantitative data, was shown by a summary circle diagram divided into five classes (size categories). Overlapping signatures were displaced using the *Convert to graphics* tool. OD matrices with and without the bridge were used to show flows between villages with the help of *XY to Line* tool [1]. With this tool, the same colour and width were assigned to all lines connecting the start and the end points of the flows with the same number of trips. The flow lines were developed thanks to the introduction of intervals for the number of journeys between villages, presented with lines of different width and brightness. An increase in line thickness and brightness as well as the colour changes suggest a higher number of trips between two locations. Using only a width variable would decrease map legibility. The same intervals for the value of the number of trips were applied for the presentations with and without the bridge ensuring the comparability of both visualizations. Thanks to the use of ArcMap software it was possible to add important map elements such as the legend, the graphical scale and the north arrow.

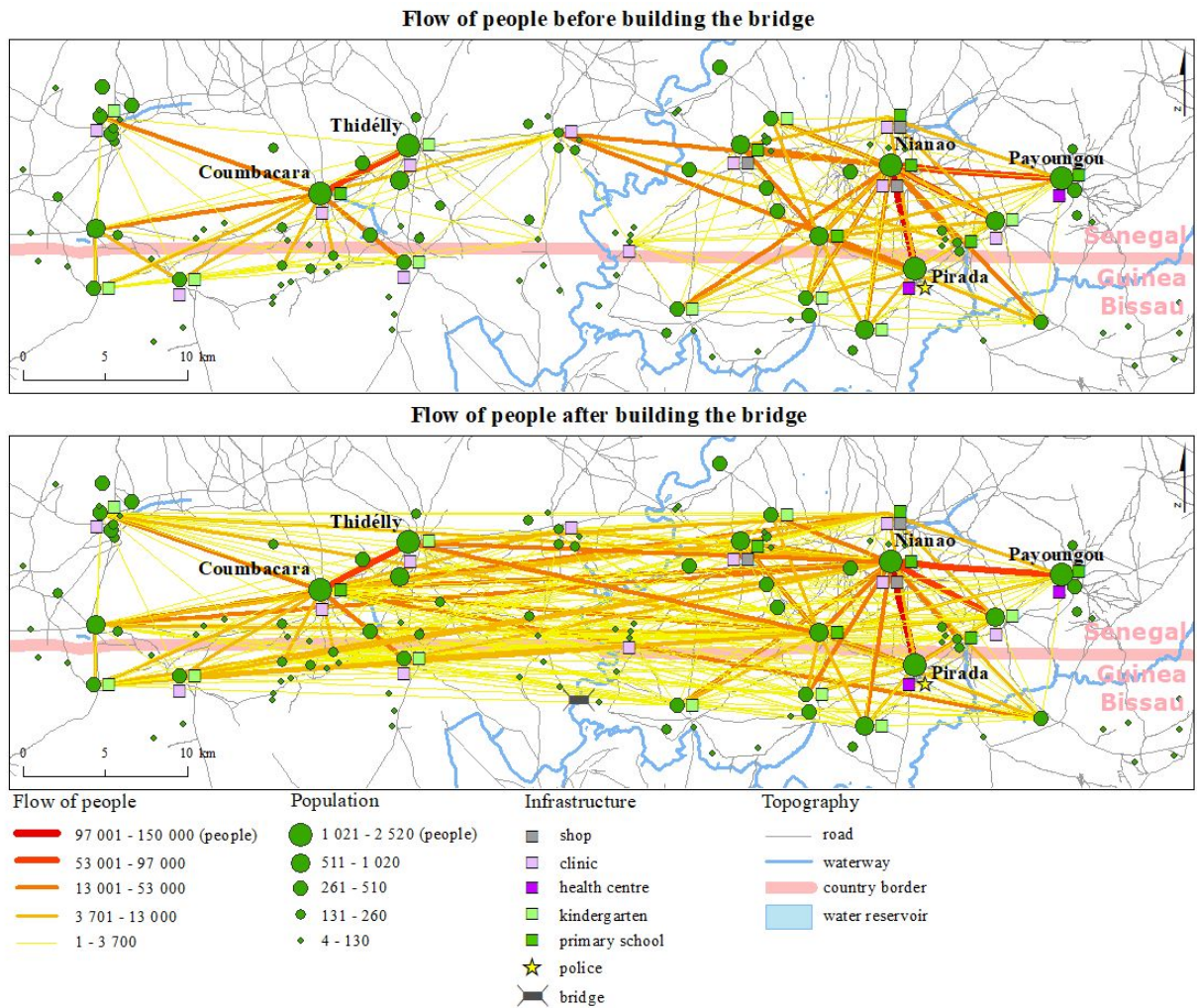

Figure 8: Maps from ArcMap: Flow of people without the bridge - visualization of tables 2,3; Flow of people with the bridge - visualization of tables 4-7.

The visualizations prepared with JFlowMap show flows of people between villages using flow lines with different width and colour at the start and the end points, which helps to recognize the direction of the journeys (Figure 9 and 10). In this software, the volume of flows between two locations is also reflected in the point signature for each locality through the use of the summary circle diagrams of different sizes. This helps to recognize where villages are located and the relevance of a given village in terms of trip attraction.

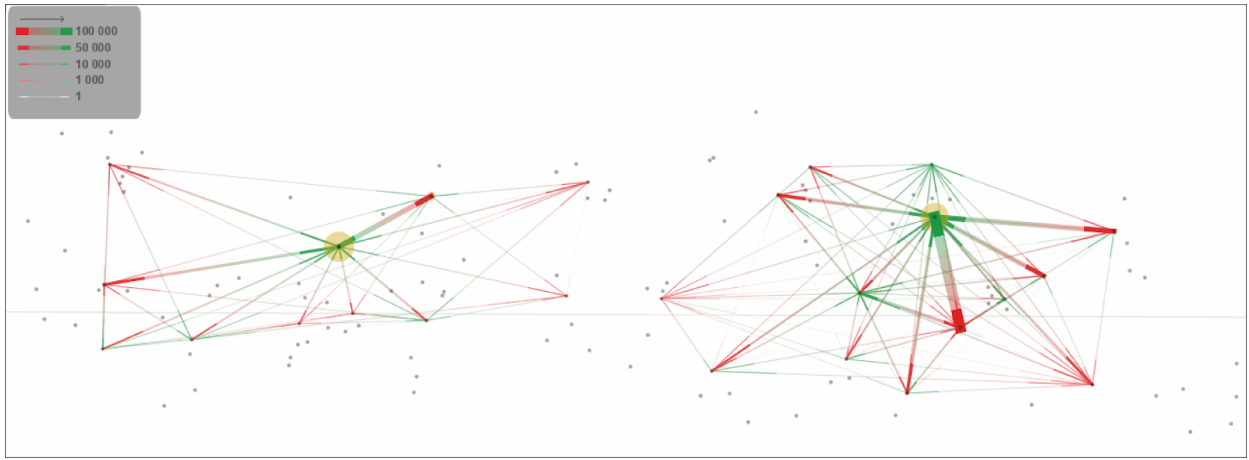
Figure 9: People flow without the bridge, prepared in JFlowMap (visualization of Tables 2 and 3).

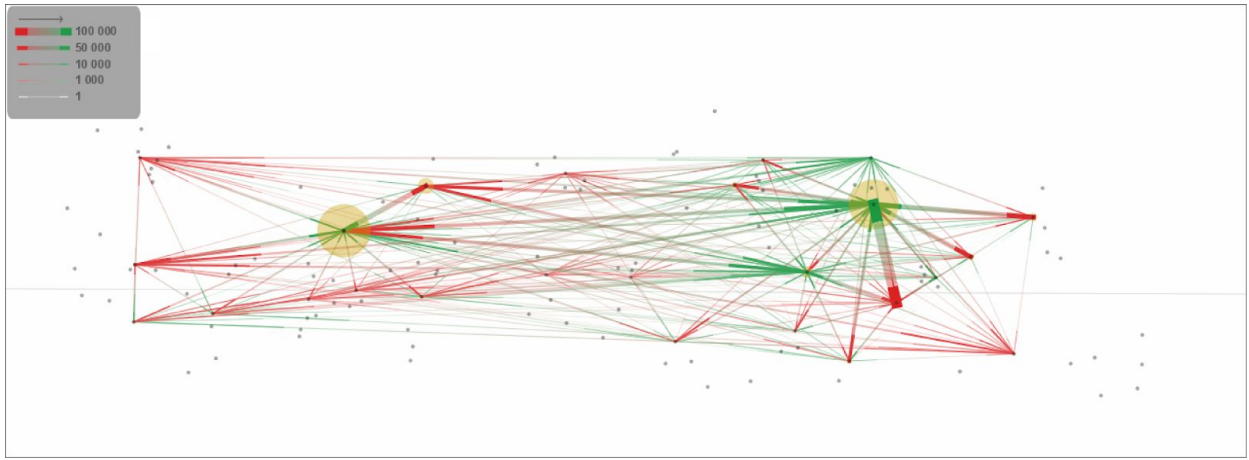
Figure 10: People flow after building the bridge, prepared in JFlowMap (visualization of Tables 4 to 7).

### 6.3 Results' analysis.

As the maps included in Section 6.2 show, when the possibility to cross the river exist, one of the main flows is the one that connects a major town on the west side of the river (Coumbacara) with the major town on the East (Nianao). This finding could be taken as an argument to support the need for the construction of a crossing infrastructure on the Kayanga-Geba river. However, without a clear distribution of trips among roads, it is impossible to determine the optimal location of the bridge: ideally on the most transited road.

Regarding the direction of the movement, it seems to be predominantly from West to East (74 per cent of the trips, see Section 6.1) and from South to North (93 per cent of the trips, see Section 6.1), with two dominant poles (Nianao and Coumbacara), both located in Senegal, at both sides of the river. The West-to-East trend could be explained by the location of markets mainly in the east side[9], since no specific maps were generated for each infrastructure, no further in-depth analysis can be done. Regarding the North-to-South, the results obtained in the mobility visualization seem to partially challenge the *border asymmetries* hypothesis as exposed in Section 4.1. A priori, Senegal holds a greater level of development than Guinea-Bissau, so it would be logical to assume better infrastructure and services to be present on this side of the border, acting as an attraction pool for Guineans; a hypothesis that results seem to support based on the regional relevance of Nianao and Coumbacara. However, the results obtained in Figures 9 and 10, although do represent a higher attraction of the Senegal side, show a more bidirectional network of cross-border mobility. A result that aligns with *Alianza*'s observation, which sees in the partial abandonment of the Casamance region, far away from capitals' networks of development (both, at national - Dakar and Bissau - and regional - Kolda and Gabu or Bafata - levels), a possible explanation. This result could help to construct the supplementary hypothesis that, in cases similar to the one analyzed in this study, the rural and remote nature of the area has also an effect on shaping the direction of mobility patterns, together with the development gap between countries: in areas where infrastructure and services are so scarce, any infrastructure becomes a pole of attraction regardless of the country it belongs to. Therefore, it could be argued that, on top of *border asymmetries*, rural hurdles to development should also be taken into account when building assumptions and analyzing mobility patterns in border-areas.

It would be reasonable to argue that the above supplementary hypotheses to *border asymmetries* is difficult to sustain with "only" a 7 percent of South-to-North, less-to-more developed, trips. Nonetheless, as previously mentioned, it was not possible to include social ties, cultural and family relations, into the study due to the difficulties to translate them to quantitative data. These ties, as indicated in Section 1 and observed by *Alianza*, are central to cross-border mobility; therefore, their exclusion from the study could have significant repercussions on the results. Repercussions of especial relevance when linked to the notion of *border asymmetries* since, in *Alianza*'s experience, social mobility drivers do not obey economic notions of development, thus transcending this notion. With the information at stake it is impossible to adventure an hypothesis on the quantity or direction of socially driven mobility vis à vis economic one (the one based on infrastructure as analyzed in this paper). Nonetheless, it raises a point of concern about the suitability of *border asymmetries* and development gaps to explain cross-border mobility patterns, perhaps too rooted on economic, technocratic approaches to development, leaving outside relevant social perspectives.

In relation to the comparison between hypotheses, without (current situation) and with (future hypothesis) the bridge, in the event that this infrastructure exists, we observe an increase in the overall number of trips: a 17 percent increase from 1,371,283 without to 1,657,493 trips with bridge as per de matrices presented in Section 6.1. This increase could indicate that the potential gain in cross-river trips does not seem to be currently covered through "inter-area" trips. A possible explanation for the former could come from the understanding of the mobility modelling: trips are the result of the connection of attraction and production poles, when one of them is left out of the study area, no trips are generated. In this case the "being left out" can be read as a synonym of inaccessibility: the attraction pole (one of the analyzed facilities) is currently inaccessible without the bridge, and there is not a substitute within the remaining area (otherwise the number of "intra-area" trips in the "without bridge" scenario would match the "with bridge" amount). Therefore, it could be hypothesized that regional accessibility, and therefore use[10], to the different facilities analyzed has the potential to be improved by the presence of the infrastructure. Since markets are the infrastructures only present on one of the sides of the river, it would be logical to assume that commerce will be the activity benefiting the most from the rise in cross-river accessibility.

---

[9] An observation shared by *Alianza* that supports this hypothesis is the presence of a weekly market in Diaobe, town located at the East of Nianao, outside of the boundaries of this study. Diaobe´s market is the most important sub-regional market of the area, destination of goods from the Gambia, Mali, Guinea Conakry, Burkina Faso and Guinea Bissau.

Two collateral comments are important to be made in relation to the above paragraph, both potentially compromising the accessibility potential of infrastructure development. Firstly, the hypothesis of absolute isolation between sub-areas should be seen as a simplification, since cross-river mobility exists even without the existence of the bridge. Secondly, it should be noticed that the notion of increased accessibility rests on the assumptions of unlimited service capacity and commuting ability. Further analysis is needed on the capacity of currently present infrastructures to effectively respond to citizens' current and potential demand, based on which a limit should be introduced to the model. Similarly, a cap to the distance individuals are able to cover by the different transport means available should be established within the model.

Finally, in relation to the border, we see that under the assumption of no-limitation to people's movement, cross-border trips are well present in the area: cross-border trips account for 31 percent of the trips obtained through the mobility model (see matrices from section 6.1).

# 7. Conclusions

### 7.1 Research limitations and areas for further research

As already mentioned, data limitations were one of the main constraints to the study. The accuracy of the results has been questioned throughout the research and some big (but reasonable) assumptions were unavoidable due to the lack of disaggregated data for the study area. However, it is important to remember that the study behind this paper was built on one of the first efforts to understand the needs and dynamics of the cross-border population in West Africa, the SAGE Program and the PAGET plans. With this in mind, data scarcity, rather than a disqualifying limitation of the results, becomes an unavoidable context and a claim for further research on cross-border mobility as a means to ensure inhabitant's wellbeing. The following paragraphs summarize the main limitations derived from the lack of data.

During the construction of the OD matrices, not enough disaggregated data was available in relation to local use of health, education, commerce facilities, modes of transports, mobility drivers and social demographics. Reasonable assumptions were made based on regional or national data, compromising, nonetheless, the accuracy of the results.

During the visualization, the lack of a georeferenced road network also impacted the analysis: without the possibility to distribute trips among the different roads, it was not possible to determine the optimal location for the bridge, only the idea that a crossing infrastructure over the river is needed, due to the dominance of cross-river trips, could be implied. Including the road network into the mobility visualizations would also have made the results more accurate in terms of adjusting the visualizations to the terrain topography.

---

[10] It has been assumed that increased access would unquestionably and effectively translate into higher use. This assumption could be questioned and further research should be conducted before the claim can be made. Nonetheless, such research falls out of the scope of this paper.

Also related to the lack of disaggregated data. Although qualitative efforts were made by *Alianza* to include a gender perspective during the elaboration of the PAGET Plans, the lack of quantitative disaggregated data by gender made impossible to generate specific maps to analyze possible cross-mobility disparities between genders.

In addition to data limitations, some decisions made during the study process also compromised the usability of the results obtained.

Although social ties seem to be one of the main drivers of cross-border mobility in the area, the constraints to translate qualitative notions into quantitative terms due to the lack of research and literature on the topic, made impossible to quantify, visualize and analyze social influence on cross-border mobility.

During the analysis phase (Section 6.3), it became clear that a constraint in the conception of the study was the decision to generate only mobility-aggregated maps instead of preparing separate maps concerning the relation between a particular infrastructure group (health, education and commerce) and peoples' mobility. Due to the above mentioned data limitations during the modelling process, and the derived qualitative partial inaccuracy of the results, the generation of individual visualizations for each type of infrastructure would have had higher qualitative value as it would have allowed to more easily analyze flow dynamics particular to the different infrastructure groups: mobility patterns and directions.

In terms of the technical aspects of the designed visualizations, it would have been beneficial to further develop some dedicated tools, especially in the JFlowMap program, making possible to add more cartographic elements to the maps (for instance the graphical scale or more informative thematic legend, geographical names, background map). These elements are available in the ArcMap program. Among the disadvantages of visualizations prepared with JFlowMap the following ones can be also named restrictions on map preparation (visualization without graphical scale) and the need to transform input data to the appropriate format. However, the big advantage of this software is the possibility of getting more information about flows using designed maps in the computer environment. The user can quickly change the value ranges and easily read information about flows for individual villages or connections.

The biggest constraint of both considered programs, influencing map legibility and comprehensiveness, constitutes the cases when the flows overlap with each other or with other background linear data. Then, the detailed map analysis may be complicated as it is impossible to recognize the precise number of flows. Thus, the flow presentation should have the priority over other background layers in the map and flow lines should be ordered in an incremental manner. Additionally, if we access the tabular data prepared in ArcMap or checked the flow value directly in the JFlowMap interface, further in-depth data analysis is possible. Despite the mentioned limitations, the results show that it is possible to recognize the general characteristics of the studied mobility patterns. This proves the added value of graphic representation as a complement to tabular representation. The solution to this issue may be the use of more advanced cartographic presentation methods which gives the possibility to interactively observe dynamic flow changes. For this purpose interactive or animated maps could be designed, for instance the visualization of cycling mobility [56].

Hopefully, the above-outlined limitations can serve as a solid ground for future, better-informed, studies. In addition, despite the inaccuracies, we believe the work done can already help to make some conclusions on the research questions indicated in Section 3.4, as summarized in the following sections.

## 7.2 Infrastructure and local transnational development

An interesting finding of this research is the increase in the number of trips, cross-river and overall, when the possibility of crossing the river is assumed; presumably due to an increase in accessibility to, and therefore use of, the different infrastructures analyzed, mainly markets. This observation might have a worth mentioning consequence when linked to governance and development. A positive correlation seems to be present between hard infrastructure development and inhabitants' accessibility and use of the soft infrastructure available within the study area, arguably with positive repercussions on their well-being. Therefore, under the assumption of the comparative cost of building hard infrastructure (i.e., bridge) being lower than that of duplicating soft infrastructure (i.e., education and health facilities), as well as the assumptions of unlimited service capacity and ability to commute (see section 6.3), facilitating cross-border mobility[11] can be seen as a governance strategy with positive effects on citizens' development and well-being. Such a hypothesis could be used as an argument in favor of transborder governance cooperation as a means to guarantee free and safe movement across borders.

## 7.3 Contribution to the literature on cross-border mobility

The decision not to generate individual visualizations for each infrastructure analyzed (schools, markets and hospitals) makes it difficult to understand the effects that each of them has on the transborder mobility. However, a possible contribution could be our decision, informed by participatory qualitative analysis, to include education as a contributing factor to cross-border mobility. This is relevant because not much research was found on this topic during the review of existing literature: most of the information found made reference to higher university education and to a scale bigger than borderland areas [29, 59]. Therefore, further analysis of education-driven, transborder mobility could contribute to closing the, apparently existing, knowledge gap around the effects that primary and secondary education might have on the mobility patterns of those, especially children, living on borderland rural areas.

Additionally, the North-to-South dominant mobility trend (Figures 9 and 10) could be interpreted as a confirmation of the existing *infrastructure asymmetries* between Senegal and Guinea-Bissau. Under this interpretation, that the main attraction poles are located in Senegal is of no surprise since better quality infrastructure is expected to be in the country with the higher development index. However, Guinea-Bissau seems to also act as an attraction pole. This makes us question the notion of *asymmetry* in the context of border areas. An assumption that, although logical, looks as a simplification of a much more complex reality, where rural isolation from national development networks, as well as social drivers, might also have an influence on the direction of and relevance of transborder mobility. The risks to go on without challenging the former assumption are three-fold: (1) to neglect the issues of limited resources available to rural populations living in border areas, normally outside of national development agendas, something no country seems to have overcome regardless of their development index; (2) to hamper cross-border cooperation and governance when contributions and beneficiaries seem to be coming, exclusively, from different sides of a border; (3) to adopt an excessively infrastructure-based, technocratic approach to development, neglecting social components that do not respond to, transcend and add to economic notions of linear and unidirectional development.

---

[11] Conclusion made based on the analysis of a geographical border (river) rather than on a political one. They are assumed as interchangeable to be able to relate to most of the existing literature, and sections of this paper, where mainly political borders are considered.

## 7.4 Visualization and governance

Departing from Raeymaekers' conceptualization of borderland practices as crucial to negotiate the constitution of power between state and non-state actor, national and local ones over these peripheral territories (see section 4.1), with the present mobility study we aimed to understand how to support local, informal transborder land governance structures to prove their legitimacy and capacity against, or as complement to, national administrations. So far, through the analysis of the obtained mobility patterns, we were able to build some hypotheses in relation to the main attraction poles, directions of movement, as well as to prove the relevance of cross-border mobility and begin to reflect about their main drivers. Additionally, if the linkages with the road network were carefully elaborated, it would allow to understand the main roads sustaining the movement and potential needs for improvement or enlargement. Similarly, if the service capacity of the different infrastructures (health, education and commerce) was examined against current or potential demands, requirements for new services or capacity enlargement could be identified. Nonetheless, in order to develop a visualization tool able to truly and effectively support in the prevention and arbitration of conflicts emerging from a shared transnational use of scarce resources and services, new approaches should be taken into account.

Still in relation to mobility, studying accessibility as an alternative, or complement, to mobility patterns, could be of greater relevance or immediate application to local authorities for future decision making-processes, to determine which are the most isolated or underserved areas, where development efforts should focus on. Mathon et al. [38] accessibility study at the Haitinian-Dominican border could serve as a good reference to build on.

## 7.5 Local realities and global technologies

Specially in cases like this one, where data is so scarce and computer-based methodologies not easily adaptable, qualitative local knowledge becomes a great complement and source of data for quantitative modelling, analysis, as well as finding's validation. The former speaks in favor of local communities involvement when methodologies are applied to new contexts in order to avoid important oversights in the inclusion of context-relevant data or perspective, uncommon in the cases where the methodology was originally developed.

Sustainability and relevance are two additional angles to consider when introducing new methodologies and technology. Ideally, the local population should be treated as sources of information as well as end users/receivers of the generated results [21]. Therefore, issues like the complexity, adaptability or replicability of the model should be taken into consideration, ensuring a compatibility with local resources and capacities (current or potential). In contexts like the one presented in this paper, where access to technologies, data, economic resources and human skills are so scarce, priority should be given to tools developed following needs-based, minimum necessary data and creative commons principles.

# Acknowledgements

This study was born out of a partnership for local development between the Spanish NGO *Alianza por la Solidaridad* (Alianza), in representation of local authorities, and the Civil Engineering School of the *Polytechnic University of Madrid* (UPM). The study was conducted by students and professors of the UPM and the Faculty of Geography and Regional Studies of the *University of Warsaw,* in collaboration with *Alianza* and the local NGOs FODDE (Senegal) and APRODEL (Guinea Bissau). It builds on the collected data through the SAGE Program (Food Security and Environmental Governance), financed by the Spanish Agency for International Development Cooperation (AECID).

# References

[1] ArcGIS XY to Line https://pro.arcgis.com/en/pro-app/tool-reference/data-management/xy-to-line.htm [last access: 3.12.2020]
[2] Alcorn J.B., 2000, Borders, Rules and Governance: Mapping to catalyse changes in policy and management. Gatekeeper Series no.SA91. IIED.
[3] Andrienko G., Andrienko N., 2008, Spatio-temporal aggregation for Visual analysis of movement. Proceedings of IEEE Symposium on Visual Analytics Science and Technology, Waszyngton, DC: IEEE Computer Society Press, pp. 51-58. DOI: 10.1109/VAST.2008.4677356.
[4] Bertin J., 1967. Sémiologie Graphique. Les diagrammes, les réseaux, les cartes.
[5] Bishop I.D, Lange E., 2005, Visualization in Landscape and Environmental Planning –Technology and Application. London/New York: Taylor & Francis.
[6] Bochaton A., 2011. La construction de l'espace transfrontalier lao-thaïlandais: une analyse à travers le recours aux soins. DOI: 10.4000/eps.4542
[7] Bochaton A., 2015. Cross-border mobility and social networks: Laotians seeking medical treatment along the Thai border. Social Science & Medicine, 124, pp. 364-373, DOI: 10.1016/j.socscimed.2014.10.022.
[8] Thematic mapping: http://thematicmapping.org/downloads/world_borders.php [last access: 18.12.2020]
[9] Boyandin I., 2013, Visualization of temporal Origin-Destination data. Thesis. Fryburg: UniPrint.
[10] Boyandin I., Bertini E., Lalanne D., 2010, Visualizing the World's Refugee Data with JFlowMap, Poster Abstracts at Eurographics/ IEEE-VGTC Symposium and Visualization, France.

[11] Briceño-Garmendia C., Moroz H., Rozenberg J., "Road Networks, Accessibility, and Resilience: The Cases of Colombia, Ecuador, and Peru.", 2015.

[12] Burghardt D., Duchêne C, & Mackaness W., 2014, Abstracting geographic information in a data rich world: Methodologies and applications of map generalisation. Cham, Switzerland: Springer International.

[13] Chen Y., 2015, "The Distance-Decay Function of Geographical Gravity Model: Power Law or Exponential Law?", Chaos, Solitons & Fractals, Vol. 77, pp. 174-189.

[14] Dataset from the PAGET programme, used in our experiments: https://docs.google.com/spreadsheets/d/1K9_sRC_kwCmIuU0tcKAHg03joBt8gqB1yjl6uOHKjpw/edit?usp=sharing [last access: 18.12.2020]

[15] Dent B. D., Torguson J. S., Hodler T. W., 2009, Cartography: Thematic map design, Sixth Edition. New York: McGraw-Hill.

[16] Diallo M., 2015. Mobilités socio-spatiales et production territoriale en Sénégambie. EchoGéo [En ligne], 34. URL: http://echogeo.revues.org/14411 [last access: 19.12.2020]

[17] Fabianova, J., Michalik, P., Janeková, J., & Fabian, M., "Design and evaluation of a new intersection model to minimize congestions using VISSIM software", Open Engineering 10, 2020, pp. 48-56.

[18] Fanchette S., 2001. Désengagement de l'État et recomposition d'un espace d'échange transfrontalier: la Haute-Casamance et ses voisins. Autrepart (19), 2001: pp. 91-113. DOI : 10.3917/autr.019.0091

[19] FlowMapper, https://github.com/cempro/flowmapper [last access: 3.12.2020]

[20] Geofabrik: http://download.geofabrik.de [last access: 18.12.2020]

[21] Gora P., "Designing urban areas using traffic simulations, artificial intelligence and acquiring feedback from stakeholders", Transportation Research Procedia 41, 2019, pp. 532–534. doi.org/10.1016/j.trpro.2019.09.089.

[22] Gora P., "Traffic Simulation Framework - a Cellular Automaton based tool for simulating and investigating real city traffic", in "Recent Advances in Intelligent Information Systems", 2009, pp. 641-653.

[23a] Guo D., Zhu X., 2014, Mapping Large Spatial Flow Data with Hierarchical Clustering. „Transactions in GIS" vol. 18(3), p. 421–435. DOI: 10.1111/tgis.12100.

[23b] Guo D., Zhu X., 2014, Origin-Destination Flow Data Smoothing and Mapping. IEEE Transactions on Visualization and Computer Graphics, vol. 20, no. 12, p. 2043-2052. DOI: 10.1109/TVCG.2014.2346271.

[24] Human Development Index ranking http://hdr.undp.org/en/content/2019-human-development-index-ranking [last accessed: 18.12.2020]

[25] Hong I., Jung W-S, Jo H-H, "Gravity model explained by the radiation model on a population landscape", PLoS ONE 14(6), 2019, 10.1371/journal.pone.0218028.

[26] Jenny B., Stephen D. M., Muehlenhaus J. Marston B. E., Sharma R., Zhang E., Jenny H., 2016, Design principles for origin-destination flow maps. „Cartography and Geographic Information Science", p. 1-15. DOI: 10.1080/15230406.2016.1262280.

[27] JFlowMap https://github.com/e3bo/jflowmap [last access: 3.12.2020]

[28] JFlowMap data preparation instruction https://github.com/ilyabo/jflowmap/blob/master/doc/HowToPrepareData.md [last access: 3.12.2020]

[29] Keevers L., Price O., Leask B., Dawood Sutal F. and Lim J., 2019. Practices to improve collaboration by reconfiguring boundaries in transnational education. Journal of University Teaching & Learning Practice. Vol. 16. Issue 2.

[30] Kennedy R. G., 1999, Problems of Cartographic Design in Geographic Information Systems for Transportation. "Cartographic perspectives" no. 32, p. 44-60. DOI: 10.14714/CP32.627.

[31a] Korycka-Skorupa J., 2002, Od danych do mapy. Część I. „Polski Przegląd Kartograficzny" t. 34, nr 2, s: 91-102.

[31b] Korycka-Skorupa J., 2002, Od danych do mapy. Część II. „Polski Przegląd Kartograficzny" t 34, no. 3, p.: 175-188.

[32] Kraak, M. J., 2005, Visualising spatial distributions. In P. A. Longley, & . [et al] (Eds.), Geographical information systems : principles, techniques, management and applications : abridged (pp. book 49-65, cd-rom 157-173). Hoboken: Wiley & Sons.

[33] Kumar P., Jain S.S., 2000, "Optimal rural road network planning for developing countries", Australian Road Research Bulletin (ARRB), Vol. 9, No.3, September 2000, pp 51-66.

[34] Lefebvre C., 2015. Frontières de sable, frontières de papier. Histoire de territoires et de frontières, du jihad de Sokoto à la colonisation française u Niger, XIXe-XXe siècles. Paris, Publications de la Sorbonne.

[35] Levy, C., 2013. Travel choice reframed: "deep distribution" and gender in urban transport. Environment & Urbanization, Volume 25(1), pp. 47-63. DOI: 10.1177/0956247813477810

[36] Levy, C., 2015. Routes to the just city: towards gender equality in transport planning. In: C. O. N. Moser, ed. Gender, asset accumulation and just cities. Pathways to transformation. Oxon: Routledge, pp. 135-149.

[37] Massey D., 1993. Power-geometry and a progressive sense of place, in Bird J., Curtis B., Putnam T., Robertson G., Tickner L. (ed.), Mapping the future local cultures, global change, London, Routledge: 59-69.

[38] Mathon D., Apparicio P. and Lachapelle U., 2018. Cross-border spatial accessibility of health care in the North-East Department of Haiti. DOI: 10.1186/s12942-018-0156-6

[39] McNally, Michael G., "The four step model.", 2008.

[40] NGO Alianza por la Solidaridad https://reliefweb.int/organization/aps [last accessed: 18.12.2021].

[41] "Pour une meilleur approche régionale du développement en Afrique de l'Ouest", OECD Library, https://www.oecd-ilibrary.org/development/pour-une-meilleure-approche-regionale-du-developpement-en-afrique-de-l-ouest_9789264276277-fr [last accessed: 18.12.2020]

[42] Pasławski J., 2006, Wprowadzenie do kartografii i topografii [Introduction to cartography and topography]. Wrocław: Nowa Era.

[43] Pieniążek M., Szejgiec B., Zych M., Ajdyn A., Nowakowska G., 2014, Graficzna prezentacja danych statystycznych. Wykresy, mapy. [Graphical presentation of statistical data. Charts, maps]. GUS., Warszawa.

[44] Pieniążek M, Zych M., 2020, Statistical maps. Data visualisation methods, Statistics Poland, Warszawa.

[45] Porter, G., 2008. Transport planning in sub-Saharan Africa II: putting gender into mobility and transport planning in Africa. Progress in Development Studies, 8(3), pp. 281-289. DOI: 10.1177/146499340800800306

[46] Le Programme d'Urgence de Modernisation des Axes et Territoires frontaliers www.puma.sn [last accessed: 18.12.2020]

[47] Raeymaekers T., 2009. The central margins Congo's transborder economy and state-making in the borderlands. Danish Institute for International Studies (DIIS) Working Paper 2009:25.

[48] Robinson J., 2015. Thinking cities through elsewhere: comparative tactics for a more global urban studies. Progress in Human Geography, vol. 40(I) 3-29. DOI: 10.1177/0309132515598025

[49] Sander N., Abel G. J., Bauer R., Schmidt J., 2014, Visualising Migration Flow Data with Circular Plots. ''Working Papers'' Vienna Institute of Demography. no. 2.Thematic Cartography and Geovisualization.

[50] Schroth O., 2010, From Information to Participation. Interactive Landscape Visualization as a Tool for Collaborative Planning. Tom 6 z IRL-Bericht, Institut für Raum- und Landschaftsentwicklung, Zürich.

[51] Food Security and Environmental Governance https://www.alianzaporlasolidaridad.org/noticias/sage-un-proyecto-contra-el-hambre-y-por-la-naturaleza-que-traspasa-fronteras-coloniales-en-africa-occidental [last access: 19.12.2020]

[52] Singh A., "GIS Based Rural Road Network Planning for Developing Countries", Journal of Transportation Engineering 12(8), 2010.

[53] Slocum T. A., McMaster R. B. , Kessler F. C., Howard H.H., 2014, Thematic Cartography and Geovisualization. 3rd Edition. Pearson Education Limited.

[54] Tapia Ladino M., Liberona Concha N. and Contreras Gatica Y., 2017. El surgimiento de un territorio circulatorio en la frontera chileno-peruana: estudios de las prácticas socio-espaciales fronterizas. Revista de Geografía Norte Grande, 66, pp. 117-141. DOI: 10.4067/S0718-34022017000100008

[55] Tobler W. R., 1987, Experiments in migration mapping by computer. "The American Cartographer" vol. 14, no. 2, p. 155–163. DOI: 10.1559/152304087783875273.

[56] Visualization of urban bike mobility https://uclab.fh-potsdam.de/cf [last access: 4.12.2020]

[57] World Health Organization (WHO), https://www.who.int [last access: 3.12.2020]

[58] Xiao N. and Chun Y., 2009, Visualizing migration flows using kriskograms. "Cartography and Geographic Information Science" vol. 36, no. 2, p. 183–191. DOI: 10.1559/152304009788188763.

[59] Zeleza P., 2005. Transnational education and African universities. Journal of Higher Education in Africa. Vol. 3, No. 1, pp. 1-28.